\documentclass[preprintnumbers,aps,pra,twocolumn,groupedaddress,showkeys]{revtex4}

\usepackage{amsmath}
\usepackage{bm}
\usepackage{color}
\usepackage{hyperref}
\usepackage{graphicx}
\usepackage{amsfonts}

\hypersetup{
	colorlinks=true,
	linkcolor=blue,
	filecolor=blue,      
	urlcolor=blue,
	citecolor=blue,
}

\begin{document}

\title{Non-Hermitian topology in a single driven-dissipative Kerr-cat qubit}

\author{Pei-Rong Han$^{1}$}
\author{Huiye Qiu$^{1}$}
\author{Hao-Long Zhang$^{2}$}
\author{Wen Ning$^{2}$}
\author{Zhen-Biao Yang$^{2}$}
\author{Shi-Biao Zheng$^{2}$}
\thanks{E-mail: fjqiqo@fzu.edu.cn}
\address{$^1$School of Physics and Mechanical and Electrical Engineering, Longyan University, Longyan 364012, China\\
$^2$Department of Physics, Fuzhou University, Fuzhou 350108, China}

\begin{abstract}
The intriguing physical phenomena associated with exceptional points have established non-Hermitian physics as a frontier of modern research. Recent investigations have extended non-Hermitian physics into the fully quantum domain. However, existing studies predominantly concentrate on discrete-variable quantum systems, while non-Hermitian quantum effects in continuous-variable encoded systems remain largely unexplored.
In this work, we investigate the exceptional structure for a driven-dissipative Kerr-cat qubit, realized with a Kerr nonlinear resonator. We find that the dissipation leads to a bidirectional jump between the two basis states of the cat qubit, which is in distinct contrast with the unidirectional jump associated with normal two-level systems. The competition between this jump and a single-photon drive gives rise to the emergence of third-order Liouvillian exceptional points (LEP3s), each of which corresponds to a crossing point of two lines of LEP2s.
Crucially, the single-photon drive is essential for generating the observed nontrivial Liouvillian topology, which reduces to a trivial case when the drive is turned off.
We further show that the LEP3 can exhibit the topological character of the Hamiltonian EP3s, which cannot be realized with a single qubit. 
Our work opens the possibility of realizing non-Hermitian phenomena with continuous-variable quantum systems.

\end{abstract}


\maketitle

\section{Introduction}
Non-Hermitian physics fundamentally enriches quantum dynamics through exceptional points (EPs)—singular degeneracies where eigenvalues and eigenvectors simultaneously coalesce. 
A central theme in contemporary studies lies in exploring the topological aspects of non-Hermitian systems, which lead to novel phenomena beyond the Hermitian paradigm \cite{1,2,3,4,5,6,7,8,9,10,11,12,13,14,15,16,17,18}. Recent advances have realized EPs in the full quantum regime, primarily through two frameworks: measurement-conditioned non-Hermitian dynamics via post-selection \cite{17,18,19,20,21,22,23} and intrinsic Liouvillian dynamics incorporating all quantum jumps \cite{24,25,26,27,28,29,30,31,32,33,34,35,36,37}. 
EPs arising in a non-Hermitian Hamiltonian are termed Hamiltonian EPs (HEPs), while those emerging in a Liouvillian superoperator are termed Liouvillian EPs (LEPs). 
For an $n$-dimension non-Hermitian system, the Hamiltonian is expressed as an $n\times n$ matrix, while the Liouvillian superoperator corresponds to an $n^2\times n^2$ matrix. This implies that the number of the eigenvalues of the Liouvillian superoperator is $n$ times larger than that for the corresponding non-Hermitian Hamiltonian \cite{29}.
Although the Liouvillian eigenspectrum can exhibit significantly richer exceptional structures than its non-Hermitian Hamiltonian counterpart, its unique topological features remain largely unexplored.
Moreover, prior investigations of LEPs have been mainly restricted to discrete-variable-encoded qubits or qudits \cite{24,25,26,27,28,29,30,31,32,33,34,35,36,37}. The problem remains of how to construct LEPs for continuous-variable-encoded qubits.

\begin{figure}[htbp]
	\centering
	\includegraphics[width=3.4in]{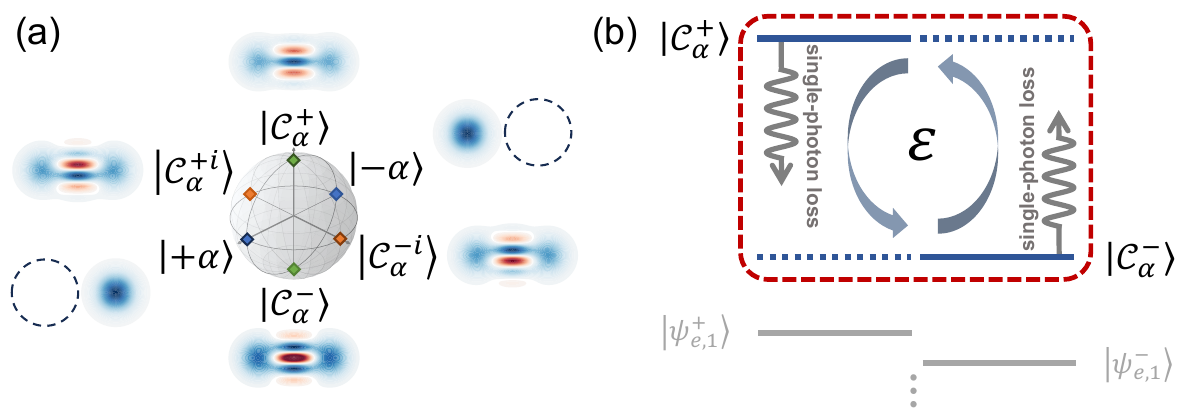}
	\caption{Kerr-cat qubit representation and effects of single-photon loss. 
(a) Bloch sphere representation of the protected Kerr-cat qubit in the large-$\alpha$ limit. Colored markers indicate the cardinal points corresponding to the encoded states, alongside their respective Wigner functions. The Z-basis states are defined as $|\pm Z\rangle=|\mathcal{C}_{\alpha}^{\pm}\rangle=\mathcal{N}_{\alpha}^{\pm}(|\alpha\rangle\pm|-\alpha\rangle)$, and the Y-basis states as $|\pm Y\rangle=|\mathcal{C}_{\alpha}^{\mp i}\rangle=(|\mathcal{C}_{\alpha}^{+}\rangle\pm i|\mathcal{C}_{\alpha}^{-}\rangle)/\sqrt{2}$.
(b) The dominant source of noise in the resonator is single-photon loss, resulting from coupling to a bath via single-photon exchange. This dissipation induces random bit flips between the states $|\mathcal{C}_{\alpha}^{\pm}\rangle$, as described by $a=\alpha(p|\mathcal{C}_{\alpha}^-\rangle\langle\mathcal{C}_{\alpha}^+|+p^{-1}|\mathcal{C}_{\alpha}^+\rangle\langle\mathcal{C}_{\alpha}^-|)$.}
    \label{Fig1}
\end{figure}

In this paper, we investigate the exceptional structure associated with the Liouvillian superoperator for a driven-dissipative cat qubit, which represents one typical example of continuous-variable-encoded qubits \cite{38,39,40,41,42,43,44}. The codewords of such a qubit is encoded in the even and odd cat states, defined as $|\mathcal{C}_{\alpha}^{\pm}\rangle=\mathcal{N}_{\alpha}^{\pm}(|\alpha\rangle\pm|-\alpha\rangle)$, where $|\pm\alpha\rangle$ are coherent states and $\mathcal{N}_{\alpha}^{\pm}=1/\sqrt{2(1\pm e^{-2|\alpha|^2}})$. These two cat states are degenerate eigenstates of a Kerr-nonlinear resonator under two-photon driving.
This qubit architecture achieves exponential suppression of bit-flip errors, positioning it as a promising paradigm for fault-tolerant quantum computation.
Within the Kerr-cat qubit framework, we encode the computational basis (Z-basis) using the parity-symmetric cat states $|\mathcal{C}_{\alpha}^{+}\rangle$ (even parity) and $|\mathcal{C}_{\alpha}^{-}\rangle$ (odd parity). The $x$-axis rotations can be implemented via single-photon driving $H_{1} = \varepsilon(a^\dagger + a)$, where $\varepsilon$ (real-valued) denotes the tunable driving strength. For large $\alpha$, the coherent states $|\pm\alpha\rangle$ serve as an approximate X-basis. Figure~\ref{Fig1}(a) depicts the corresponding Bloch sphere representation of this encoded qubit.

The Hermiticity of this system is manifested by the single-photon loss, and the transition between the even and odd cat states is driven by a classical field (Fig.~\ref{Fig1}(b)). Under the competition between the dissipation and drive, we find that the system exhibits rich exceptional structures, including second-order Liouvillian exceptional lines (LEP2 lines) and third-order LEPs (LEP3s). 
We further characterize the topological nature of each LEP3 by computing the winding number of the resultant vector \cite{45}.

\section{Theoretical framework and model}
Under single-photon loss with dissipation rate $\kappa$, the system's evolution is governed by the Lindblad master equation
\begin{equation}
    \dot{\rho} \equiv \mathcal{L}(\rho) = -i [H,\rho] + \kappa(a\rho a^{\dagger}-\frac{1}{2}a^{\dagger}a\rho-\frac{1}{2}\rho a^{\dagger}a),
    \label{master eq}
\end{equation}
where $\rho$ is the density operator, and $a^{\dagger}$($a$) is the creation (annihilation) operator for the Kerr-nonlinear resonator.
The dynamics can be fully captured by the Liouvillian superoperator $\mathcal{L}$. 
The model described here fits within the topological classification framework for driven-dissipative systems, where steady-state structures and fluctuation dynamics are characterized by graph invariants \cite{46,47}.
In the frame rotating at the driving frequency,
\begin{equation}
    H = \Delta a^{\dagger} a - Ka^{\dagger2} a^2 + P\left(a^{\dagger2}+a^2 \right) +\varepsilon(a^{\dagger}+a),
      \label{H}
\end{equation}
where $\Delta$ is the detuning between the drive and resonator, $K$ is the Kerr coefficient and $P$ is the amplitude of the two-photon driving.
For simplicity and without loss of generality, we restrict both $K$ and $P$ to be positive and real throughout this paper, which implies that $\alpha$ is likewise real.
When $\Delta =\varepsilon= 0$, the coherent states $|\pm\alpha\rangle$ are two degenerate eigenstates of the system Hamiltonian with the eigenvalue $P^2/K$, where $\alpha = \sqrt{P/K}$. 
In such a degenerate subspace, any state can also be expressed in terms of the even and odd cat states $|\mathcal{C}_{\alpha}^{\pm}\rangle$, which are orthogonal to each other and thus form the basis for a qubit. The energy gap between these degenerate eigenstates and their nearest eigenstates is $4K|\alpha|^2$. When the single-photon driving strength, dissipative rate, and detuning are much smaller than this gap, the system is approximately restricted in the subspace of the cat qubit.

\begin{figure}[htbp]
	\centering
	\includegraphics[width=3.4in]{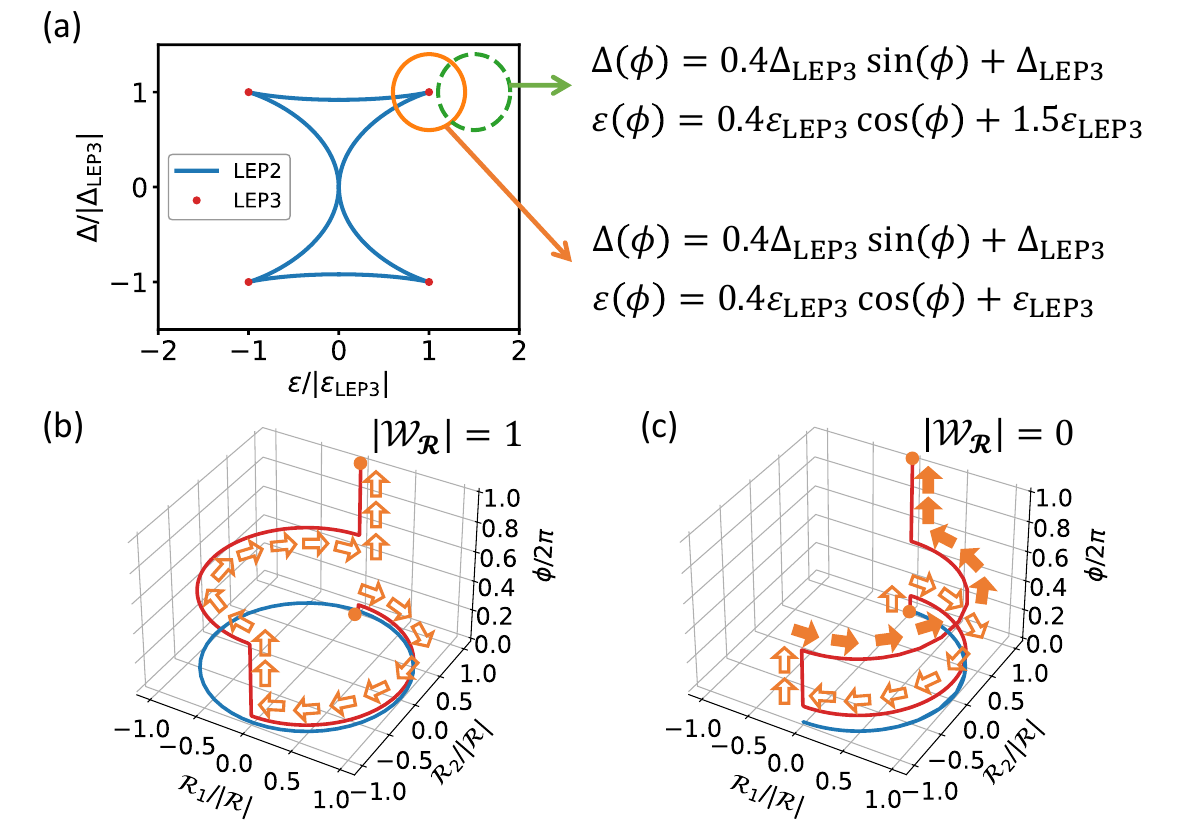}
	\caption{Topology of the Liouvillian superoperator.
    (a) Solid curves depict second-order Liouvillian exceptional points (LEP2s), along which pairs of eigenvectors coalesce. Red dots indicate third-order exceptional points (LEP3s). The solid orange circle (enclosing a single LEP3) and dashed green circle (excluding LEP3s) indicate Winding number calculation contours in parameter space. (b)(c) Winding number of the resultant vector. (b) Trajectory of the normalized resultant vector $\mathcal{R}_{\rm N} \equiv (\mathcal{R}_1 + i\mathcal{R}_2)/|\bm{\mathcal{R}}|$ under parameter variation along $\mathcal{C}_\phi$. 
    The $x$, $y$ and $z$ axes represent $\mathcal{R}_1/|\bm{\mathcal{R}}|$, $\mathcal{R}_2/|\bm{\mathcal{R}}|$ and $\phi/2\pi$, respectively. The red trajectory shows the evolution of $\mathcal{R}_1$ and $\mathcal{R}_2$ as $\phi$ varies. Its projection onto the rescaled $\mathcal{R}_1$-$\mathcal{R}_2$ plane forms a closed circle (solid blue line), indicating a topological winding number $|\mathcal{W}_{\bm{\mathcal{R}}}| = 1$.  (c) In contrast, when the projection does not form a closed circle, the winding number is $|\mathcal{W}_{\bm{\mathcal{R}}}| = 0$. The starting point and end point are marked by solid orange dots. Solid arrows indicate counterclockwise winding, while hollow arrows indicate clockwise winding.}
    \label{Fig2}
\end{figure}

\section{Liouvillian exceptional structure}
To investigate the Liouvillian spectrum and LEPs, we employ a vectorized representation to express the Liouvillian superoperator in matrix form. In this representation, Eq.~(\ref{master eq}) takes the form $\dot{\bm{V}}(t)=\mathcal{L}_{\mathrm{matrix}}\bm{V}(t)$, where $\bm{V}$ is a vectorized density matrix and $\mathcal{L}_{\mathrm{matrix}}$ is the matrix form of the Liouvillian superoperator $\mathcal{L}$
, given by
\begin{align}
    \mathcal{L}_{\mathrm{matrix}} = &-i\left( H \otimes I - I \otimes H^{\top}\right) \\ \nonumber
    &+ \left[ \Gamma_a \otimes \Gamma_a^{*} - \frac{\Gamma_a^{\dagger}\Gamma_a\otimes I}{2} - \frac{I \otimes \Gamma_a^{\top}\Gamma_a^{*}}{2} \right],
\end{align}
where $\otimes$ represents Kronecker product operation, $\top$ represents the transpose, $*$ represents the complex conjugate and $\Gamma_a=\sqrt{\kappa}a$.
Projecting $\mathcal{L}$ into the subspace spanned by $\{|\mathcal{C}_{\alpha}^{+}\rangle,|\mathcal{C}_{\alpha}^{-}\rangle\}$ yields a $4\times4$ matrix representation,
\begin{widetext}
\begin{equation}
    \mathcal{L}_{\mathrm{matrix}} =
    \left(
    \begin{array}{cccc}
        -\alpha^2\kappa p^2 & i\alpha\varepsilon p_1^+ & -i\alpha\varepsilon p_1^+ & \alpha^2\kappa p^{-2} \\
        i\alpha\varepsilon p_1^+ & \alpha^2(-\frac{\kappa}{2}p_2^+ + i\Delta p_2^-) & \alpha^2\kappa & -i\alpha\varepsilon p_1^+ \\
        -i\alpha\varepsilon p_1^+ & \alpha^2\kappa & \alpha^2(-\frac{\kappa}{2}p_2^+ -i\Delta p_2^-) & i\alpha\varepsilon p_1^+ \\
        \alpha^2\kappa p^{2} & -i\alpha\varepsilon p_1^+ & i\alpha\varepsilon p_1^+ & -\alpha^2\kappa p^{-2} 
    \end{array}
    \right),
\end{equation}
\end{widetext}
where $p_j^{\pm}=p^{-j}\pm p^j$, with $p=\mathcal{N}_{\alpha}^{+}/\mathcal{N}_{\alpha}^{-}$. The derivation relies on the relations $a|\mathcal{C}_{\alpha}^{\pm}\rangle=\alpha p^{\pm 1}|\mathcal{C}_{\alpha}^{\mp}\rangle$ and $a^{\dagger}|\mathcal{C}_{\alpha}^{\pm}\rangle=\alpha p^{\mp 1}|\mathcal{C}_{\alpha}^{\mp}\rangle$ within the subspace.
Assuming $\mathcal{L}_{\mathrm{matrix}}$ is diagonalizable, let $E_i$ and $\bm{V}_i$ denote its eigenvalues and corresponding eigenmatrices, respectively. The time evolution of $\bm{V}(t)$  is then expressed as
\begin{equation}
    \bm{V}(t) = \sum_i c_i\text{exp}(E_it)\bm{V}_i,
    \label{Liouvillian dynamics}
\end{equation}
where $c_i$ denote expansion coefficients.
Using Cardano's method, the eigenvalues of $\mathcal{L}_{\mathrm{matrix}}$ are explicitly given by
\begin{align}
    E_1 &= 0, \\
    E_2 &= -\frac{2}{3}\kappa\alpha^2p_2^+ + (\eta_{+}+\eta_{-}) , \\
    E_3 &= -\frac{2}{3}\kappa\alpha^2p_2^+ + \left(e^{i\frac{2\pi}{3}}\eta_{+}+e^{-i\frac{2\pi}{3}}\eta_{-}\right), \\
    E_4 &= -\frac{2}{3}\kappa\alpha^2p_2^+ +\left(e^{-i\frac{2\pi}{3}}\eta_{+}+e^{i\frac{2\pi}{3}}\eta_{-}\right),
\end{align}
where $\eta_{\pm}=\sqrt[3]{q\pm\sqrt{q^2+m^3}}$ with 
\begin{align*}
    q=\frac{\alpha^4\kappa}{216}[& -\alpha^2(36\Delta^2+\kappa^2)p_6^+ + 72\varepsilon^2p_4^+ \\ 
    &+ (36\Delta^2\alpha^2+576\varepsilon^2+33\kappa^2\alpha^2)p_2^+ + 1008\varepsilon^2 ],
\end{align*}
and
\begin{align*}
    m=\frac{\alpha^2}{36}[ &\alpha^2(12\Delta^2-\kappa^2)p_4^+ + 48\varepsilon^2p_2^+ \\
    &-24\Delta^2\alpha^2+96\varepsilon^2-14\alpha^2\kappa^2 ].
\end{align*}

The eigenvalues are characterized by a vanishing $E_1$ (whose eigenvector corresponds to the steady state $\rho_{\mathrm{ss}}$), a strictly real $E_2$, and a complex conjugate pair $E_3$, $E_4$.
Single-photon driving induces a topological transition where the original LEP2 at $|\Delta| = \kappa / p_2^-$ when $\varepsilon=0$ splits into two daughter exceptional points of identical order or turns to an LEP3, as shown in Fig.~\ref{Fig2}(a). 
The competition between the coherent drive and quantum jumps between $|\mathcal{C}_{\alpha}^{+}\rangle$ and $|\mathcal{C}_{\alpha}^{-}\rangle$ induced by single photon dissipation can lead to the LEP3 in the two-dimension parameter space \cite{29}.
The resulting LEP3s appear at
\begin{align}
    |\varepsilon_{\text{LEP3}}| &= \frac{\sqrt{6}\kappa}{18}\frac{\alpha\left(p^{4} + 1\right)^{3/2}}{p \left(p^{2} + 1\right)^{2}}, \\
    |\Delta_{\text{LEP3}}| &= \frac{\sqrt{3}\kappa}{18}
    \frac{(p^4+6p^2+1)^{3/2}}{\left(p^{2} - 1\right) \left(p^{2} + 1\right)^{2}}.
\end{align}
Recently, it was predicted that HEPs can emerge in a Kerr cat qubit that is subjected to dephasing, but is free of dissipation \cite{48}, which is the dominant decoherence source of the Kerr-cat qubit realized in the experiment reported in Ref.~\cite{38}.

The resulting exceptional structure 
is similar with that revealed in Ref. \cite{29}, both featuring EP2 lines intersecting at EP3 vertices.
Crucially, their physical origins differ fundamentally.
For Kerr-cat qubits, single-photon dissipation induces bidirectional quantum jumps between even- and odd-parity cat states. This effect is analogous to a quantum jump operator represented by $\sigma_x$ in a conventional qubit. The exceptional structures illustrated in Fig.~\ref{Fig2}(a) originate from this bidirectionality of quantum jumps, whereas those reported in Ref.~\cite{29} arise from unidirectional quantum jumps from the excited to the ground state. 

\section{Winding number}
Considering their apparent similarity, these configurations exhibit analogous topological properties.
The topological invariant associated with the exceptional structure can be quantiﬁed by the winding number of the resultant vector $\bm{\mathcal{R}}$ \cite{45},
\begin{equation}
    \mathcal{W}_{\bm{\mathcal{R}}}=\frac{1}{2\pi}\oint_{\mathcal{C}_{\phi}}\frac{1}{\lVert\bm{\mathcal{R}}\rVert^2}\left( \mathcal{R}_1\frac{\partial\mathcal{R}_2}{\partial\phi}-\mathcal{R}_2\frac{\partial\mathcal{R}_1}{\partial\phi} \right)\mathrm{d}\phi,
\end{equation}
where $\mathcal{C}_{\phi}$ denotes a closed contour in $\varepsilon$-$\Delta$ parameter space and is parameterized by $\phi$. 
We define the resultant vector as $\bm{\mathcal{R}} = \mathcal{R}_1 +i \mathcal{R}_2$,  whose vanishing signals the coalescence of eigenvalues and eigenvectors. 
The resultant of two polynomials is the determinant of their Sylvester matrix, vanishing if and only if the polynomials possess a common root.
Up to a nonzero multiplicative factor, the characteristic polynomial $P(E) \equiv \det[\mathcal{L}_{\text{matrix}} - E \mathbb{I}]$ of $\mathcal{L}_{\text{matrix}}$ can be written as $P(E) = \prod_{i=1}^{4}(E - E_i)$. Since the steady state and its corresponding eigenvalue do not contribute to the formation of exceptional points, the winding number is governed solely by the remaining three eigenvalues. We therefore factor out the steady-state eigenvalue and consider the cubic polynomial formed by the other three: $\tilde{P}(E) = (E - E_2)(E - E_3)(E - E_4)$.
This leads to the following explicit expressions for the resultants $\mathcal{R}_1$ and $\mathcal{R}_2$ in terms of eigenvalues $E_i$
\begin{equation}
    \mathcal{R}_1 = - \left(E_{2} - E_{3}\right)^{2} \left(E_{2} - E_{4}\right)^{2} \left(E_{3} - E_{4}\right)^{2},
\end{equation}
and
\begin{equation}
    \mathcal{R}_2 = -8 \left(E_{2} + E_{3} - 2 E_{4}\right) \left(E_{2} + E_{4} - 2 E_{3}\right) \left(E_{3} + E_{4} - 2 E_{2}\right).
\end{equation}

As established in Ref.~\cite{45}, a value of $|\mathcal{W}_{\bm{\mathcal{R}}}| = 1$ signifies that the region enclosed by the contour contains an LEP3, where $\mathcal{R}_1 = \mathcal{R}_2 = 0$. 
Figure~\ref{Fig2}(b) and (c) illustrate the behavior of the rescaled $\mathcal{R}_1$ and $\mathcal{R}_2$ as functions of $\phi$ for two different contours $\mathcal{C}_{\phi}$: one encircling only an LEP3, and one excluding the LEP3. The results demonstrate that when an LEP3 is enclosed, $|\mathcal{W}_{\bm{\mathcal{R}}}| = 1$.
Therefore, the LEP3 shares the same topological properties as an HEP3 reported in Ref.~\cite{17}.

The winding number $\mathcal{W}_{\bm{\mathcal{R}}}$ associated with a specific loop in parameter space is obtained from the Liouvillian eigenvalues along the loop. The Liouvillian eigenvalues at each point can be inferred from the time-evolving density matrix under the Liouvillian dynamics. The density matrix can be measured by quantum state tomography techniques demonstrated in recent experiments \cite{17,35,37}.

\begin{figure}[htbp]
	\centering
	\includegraphics[width=3.4in]{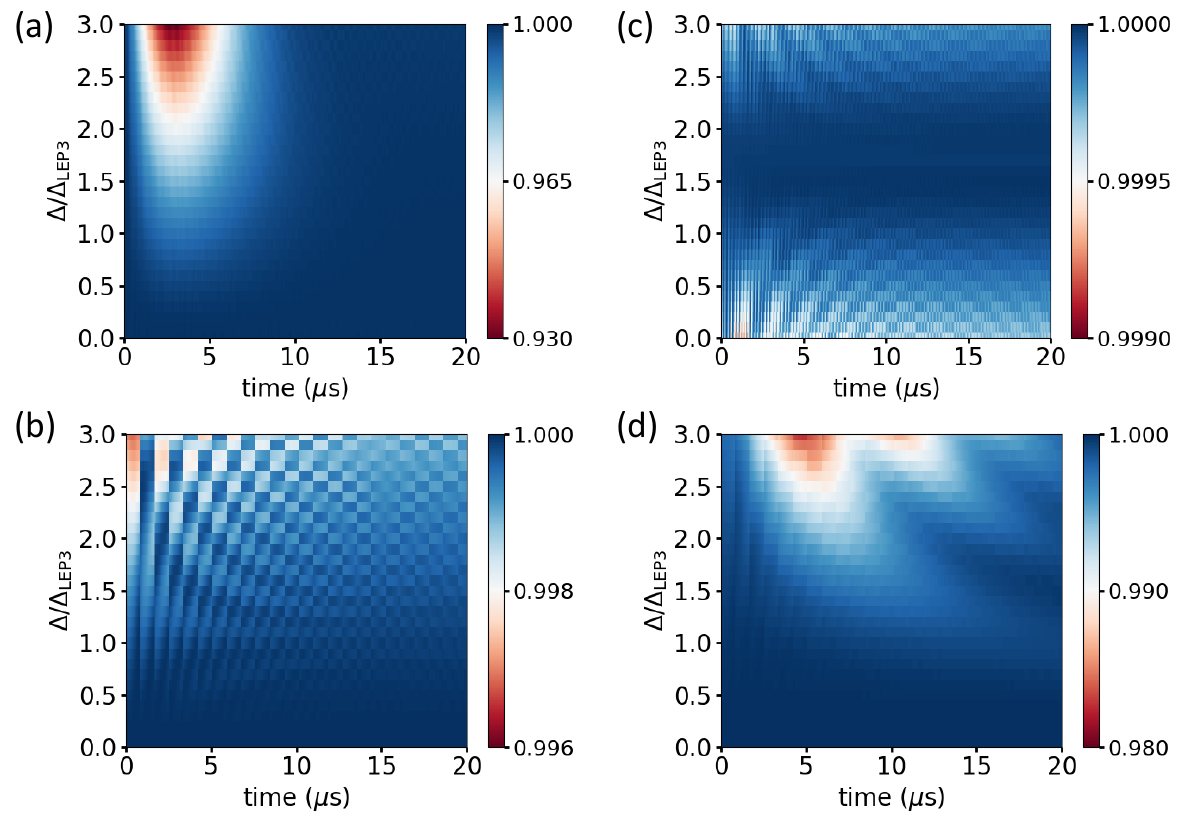}
	\caption{Validation of the Liouvillian dynamics. Color represents the fidelity $F(\rho_{\text{H}},\rho_{\text{L}})$ between states evolved under the Lindblad master equation and the Liouvillian approximation.
    (a) Initial state $|\mathcal{C}_{\alpha}^{+}\rangle$ with $\varepsilon/2\pi = 0.74$ MHz. 
    (b) Initial state $|\mathcal{C}_{\alpha}^{+}\rangle$ without single-photon driving. 
    (c) Initial state $|\alpha\rangle$ with $\varepsilon/2\pi = 0.74$ MHz. 
    (d) Initial state $|\alpha\rangle$ without single-photon driving. Numerical parameters are taken from Ref.~\cite{38}: $K/2\pi = 6.7$ MHz, $\kappa /2\pi \approx 10.3$ kHz, $|\alpha|^2 = 2.3$.}
    \label{Fig3}
\end{figure}

\section{Analysis of experimental feasibility}
It is necessary to show to what extent the system can be restricted in the subspace of the Kerr-cat qubit during the state evolution. To do so, we respectively calculate the density matrix ($\sigma$) evolved under the effective Liouvillian dynamics of Eq.~(\ref{Liouvillian dynamics}) and that ($\rho$) governed by the original master equation of Eq.~(\ref{master eq}). The validity of the approximation is quantified by the fidelity, defined as $F(\rho,\sigma)=(\text{Tr}\sqrt{\sqrt{\rho}\sigma\sqrt{\rho}})^2$. The fidelities for different initial states, as functions of the detuning and time, are presented in Fig.~\ref{Fig3}. In the simulation, we use the parameters reported in the experiment of Ref.~\cite{38}: $K/2\pi = 6.7$ MHz, $P/2\pi=15.5$ MHz, and $\kappa=1/15.5$ $\mu s^{-1}$ (i.e., $\kappa /2\pi \approx 10.3$ kHz). The resulting coherent-state amplitude is $\alpha\approx1.52$. $\varepsilon$ is set to $0.74\times 2\pi$ MHz, which corresponds to the single-photon drive strength used for implementing the X-rotation in the experiment reported in Ref.~\cite{38}.
As shown, the two dynamical evolutions exhibit excellent agreement over the selected parameter range, with fidelity exceeding 0.930. The fidelity asymptotically approaches 1 as the system relaxes to its steady state with increasing time. Minor deviations originate primarily from the detuning $\Delta a^{\dagger}a$ and the single-photon driving $\varepsilon(a^{\dagger} + a)$, which are not fully captured in the effective Liouvillian description.

To clarify the underlying mechanisms, we analyze the individual effects of the detuning and the single-photon driving.
The detuning term $\Delta a^{\dagger}a$ in Eq.~(\ref{H}) effectively 
lifts the degeneracy of the even and odd cat states. This is due to the fact that they have slightly different photon numbers, given by $\alpha^2p^2$ and $\alpha^2/p^2$, respectively. Consequently, the expectation values of this term are different for these two cat states. Furthermore, this term can induce transitions from these cat states to other eigenstates of the Hamiltonian with $\Delta=\varepsilon=0$. The single-photon driving and dissipation can also make the system jump out of the subspace spanned by the cat qubit. 
While the fidelity remains high for the experimental loss rate, significantly larger $\kappa$ would degrade the confinement to the cat manifold and reduce the accuracy of the effective model.
Increasing the single-photon loss rate by a factor of three reduces the fidelity of the Liouvillian evolution in Fig.~\ref{Fig3}(a), yielding a minimum of $\sim 0.90$.
In principle, these unwanted effects can be mitigated by decreasing the values of the parameters $\kappa$, $\Delta$, and $\varepsilon$.

In experiments, aside from single-photon loss, other decoherence channels such as two-photon loss ($\kappa_2$) and dephasing ($\kappa_{\phi}$) are also present. Two-photon loss is commonly employed to stabilize cat states. Both even- and odd-parity cat states are eigenstates of the two-photon loss operator $a^2$, satisfying $a^2|\mathcal{C}_{\alpha}^{\pm}\rangle = \alpha^2|\mathcal{C}_{\alpha}^{\pm}\rangle$. Consequently, two-photon loss does not induce transitions between the even and odd cat states and acts uniformly on all states within the cat-encoded subspace. In the cat-state basis, the two-photon loss operator is given by a diagonal matrix with equal values, and thus does not affect the dynamics of the cat qubit. Therefore, it does not contribute to $\mathcal{L}_{\text{matrix}}$ and does not alter the positions of the LEPs. When we follow Ref.~\cite{39} by setting $\kappa_2 = \kappa$, the fidelity of the Liouvillian dynamics shown in Fig.~\ref{Fig3} is further improved slightly. As for dephasing, it generally has a much weaker effect on the system compared with single-photon loss ($\kappa_{\phi} \ll \kappa$). While dephasing modifies the structure of the Liouvillian spectrum and shifts the positions of the exceptional points in parameter space \cite{37}, the resulting LEP3 remains topologically nontrivial.
The topology is robust against small parameter fluctuations. As long as the parameter fluctuations do not cause the loop not encircle the LEP3, the associated winding number remains $\pm 1$.

\section{Conclusion and outlook}
In summary, we have investigated the exceptional structure and topological properties of a Kerr-cat qubit realized in a driven-dissipative Kerr resonator. We have discovered that the LEP3 shares the same topological characteristics as an HEP3, as captured by the winding number $\mathcal{W}_{\bm{\mathcal{R}}}$ of the resultant vector. 
Using experimentally relevant parameters, our numerical simulations validate the effectiveness of the Liouvillian dynamics, confirming the experimental feasibility of the predicted exceptional structure.
The exclusive presence of LEPs indicates that they are solely induced by bidirectional quantum jumps, underscoring the fundamental distinction between LEPs and HEPs in the quantum regime. 
Notably, the nontrivial topology emerges only when the single-photon drive is present ($\varepsilon \neq 0$). The exceptional structure becomes trivial in the absence of driving, underscoring the synergy between coherent driving and dissipative jumps.
The demonstrated LEP3s in a continuous-variable qubit open several future avenues. These include braiding of exceptional points in higher-dimensional parameter spaces, enhanced sensing near third-order singularity, non-adiabatic control of quantum states using Liouvillian dynamics, and extending such topological phenomena to multimode bosonic systems or other continuous-variable encodings. These directions could further bridge continuous-variable quantum systems with non-Hermitian physics and related applications.
Our work paves the way for the exploration of exotic features in NH continuous-variable-encoded qubits.

This work was supported by the National Natural Science Foundation of China
(Grant Nos. 12505021, 12474356, 12274080, 12475015, 12505016), the
Natural Science Foundation of Fujian Province (Grant Nos. 2025J01383, 2025J01465) and the Research Startup Funds of Longyan University (LB2025002).

{\bf Conflict of Interest}  The authors declare that they have no conflict of interest.

\end{document}